\documentclass[aps,prl,a4paper,twocolumn,showpacs,superscriptaddress,floatfix]{revtex4}
\usepackage{dcolumn}
\usepackage{bm}
\usepackage{rotating}
\usepackage{lscape}
\usepackage{stmaryrd}
\usepackage{latexsym}
\usepackage{amsmath}
\usepackage{graphics}
\usepackage{graphicx}
\usepackage{bm}
\usepackage{amssymb}
\usepackage{color}
\usepackage{fancybox}
\usepackage{amsfonts}
\usepackage{afterpage}
\usepackage{epsfig}
\allowdisplaybreaks

\newcommand{\be}{\begin{equation}}
\newcommand{\ee}{\end{equation}}
\newcommand{\bea}{\begin{eqnarray}}
\newcommand{\eea}{\end{eqnarray}}
\newcommand{\ba}{\begin{eqnarray*}}
\newcommand{\ea}{\end{eqnarray*}}
\newcommand{\dagga}{{\phantom{\dagger}}}

\newcommand{\bk}{\mathbf{k}}

\newcommand{\bkp}{\mathbf{k'}}

\newcommand{\fract}[2]{\frac{\displaystyle #1}{\displaystyle #2}}

\newcommand{\eqn}[1]{(\ref{#1})}

\begin{document}

\title{Ferromagnetic Kondo Effect at Nanocontacts}
\author{Paola Gentile}
\affiliation{International School for Advanced Studies (SISSA), and CRS Democritos, CNR-INFM,
Via Beirut 2-4, I-34014 Trieste, Italy} 
\affiliation{Laboratorio Regionale SuperMat, CNR-INFM,    
and Dipartimento di Fisica ``E. R. Caianiello'', Universit\`a di Salerno, I-84081 Baronissi, Salerno, Italy} 
\author{Lorenzo De Leo}
\affiliation{Centre de Physique Th\'eorique, Ecole Polytechnique, CNRS, 91128 Palaiseau, France}
\author{Michele Fabrizio}
\affiliation{International School for Advanced Studies (SISSA), and CRS Democritos, CNR-INFM,
Via Beirut 2-4, I-34014 Trieste, Italy} 
\affiliation{The Abdus Salam International Centre for Theoretical Physics 
(ICTP), P.O.Box 586, I-34014 Trieste, Italy}
\author{Erio Tosatti}
\affiliation{International School for Advanced Studies (SISSA), and CRS Democritos, CNR-INFM,
Via Beirut 2-4, I-34014 Trieste, Italy} 
\affiliation{The Abdus Salam International Centre for Theoretical Physics 
(ICTP), P.O.Box 586, I-34014 Trieste, Italy}
\date{\today}
\begin{abstract}
Magnetic impurities bridging nanocontacts and break junctions of nearly magnetic 
metals may lead to permanent moments, analogous to the giant moments well known
in the bulk case. A numerical renormalization group (NRG) study shows that, 
contrary to mean field based expectations, a permanent moment never arises 
within an Anderson model, which invariably leads to strong Kondo screening. By including in
the model an additional ferromagnetic exchange coupling between leads and impurity, 
the NRG may instead stabilize a permanent moment through a ferromagnetic Kondo effect.
The resulting state is a rotationally invariant spin, which differs profoundly from 
mean field. A sign inversion of the zero-bias anomaly and other
spectroscopic signatures of the switch from regular to ferromagnetic Kondo are 
outlined.
\end{abstract}

\maketitle
Zero-bias anomalies that emerge ubiquitously in tunneling spectroscopy 
across quantum dots~\cite{Goldhaber-Gordon-Kondo} and single-molecules~\cite{Yu&Natelson} 
are known to be a manifestation -- in a context far away from traditional magnetic alloys -- 
of the Kondo effect~\cite{Ng&Lee,Glazman&Raikh} -- 
the phenomenon that leads magnetic impurities to lose their magnetic moments 
when diluted in a nonmagnetic metal.
Actually, Kondo screening is not an inevitable fate for magnetic impurities in 
metals. Well known counter-examples are the giant permanent magnetic moments that arise 
when $3d$ transition metal impurities are diluted in a nearly ferromagnetic metal
such as Pd~\cite{Pd-review}. We consider here the novel possibility that permanent moments 
may encounter a revival in magnetic contacts and nanoelectronics. A magnetic impurity
could act as the bridging atom in a Pd or Pt break junction~\cite{agrait}, or close-by
STM atomic contact~\cite{neelPRL2007}. Alternatively, the short monatomic chains that forms 
spontaneously in pure Pt or Ir break junctions could spontaneously and locally 
magnetize, as suggested by calculations~\cite{Delin-PRL,Smogunov-2008},
giving rise to rather similar physics to that of a magnetic alloy, even though here
the host metal (the leads) and the impurity (the nanocontact) are the same material.
In either case, the permanent moment will avert Kondo screening by the leads, 
and the standard zero bias anomaly picture does not apply. On the other hand,
one cannot expect like in a ferromagnet two different conductances for spin 
up and for spin down, because in zero field there is no up and no down, and
conductance must be independent of the lead electron spin. Sophisticated many-body 
techniques have been specifically designed during the years to study the regular 
Kondo effect~\cite{Hewson}. On the contrary, the reverse phenomenon of permanent 
moment formation has generally been dismissed after a simple treatments such as 
Hartree Fock (HF) or density functional theory (DFT), whose mean field nature is 
bound, as we will show, to miss important many-body effects. Actually,
for the giant moments generated by magnetic impurities in alloys, the crude static mean 
field description is in the end justified by finite concentration, 
giving rise to RKKY-mediated long range ferromagnetic order 
at low temperature~\cite{Pd-review}. 
No such justification is possible for nanocontacts, where quantum fluctuations and 
many-body effects must play an essential role, since local magnetism cannot 
induce bulk magnetism in the leads. 

In this Letter we examine a magnetic impurity in a nearly ferromagnetic
nanocontact (and re-examine that in a bulk host metal too) through an effective 
Anderson impurity model with 
parameter values that would suggest ferro coupling and permanent moment formation in mean field. 
We find that inclusion of many-body effects, accomplished by means of the 
numerical renormalization group (NRG)~\cite{Krishnamurthy-2}, 
alters drastically the mean-field result, and invariably leads to regular 
Kondo screening, so that no permanent local moment can survive. As NRG shows,
only through addition to the bare
Anderson model of a direct {\sl intersite} ferromagnetic exchange 
coupling between the magnetic impurity and the leads a permanent impurity 
spin can be stabilized. In this case the regular (antiferromagnetic) Kondo effect
and the associated screening with formation of a spin singlet is turned 
over to a {\it ferromagnetic} Kondo effect, with {\it anti}screening and formation
of a fully rotationally invariant, integer or half integer spin $S$. 
This role of direct impurity-lead ferromagnetic exchange, consistent with the impurity spin dynamics measured 
by perturbed $\gamma$-ray distribution in bulk alloys~\cite{Riegel,Mishra}, could be crucial at nanocontacts
in nearly ferromagnetic metals, where spectroscopic signatures of the ferromagnetic 
Kondo effect may be observable in $I$-$V$ conductance anomalies.  
      
Realistic density-functional theory local spin-density (LSDA) calculations~\cite{Zeller-PRB}
have long shown that, while earlier transition metal elements impurities like Cr tend to 
counter-polarize a bulk host metal like Pd, later elements, like Fe and Co, 
do the reverse, and co-polarize Pd, in accord with the absence/presence of permanent moments 
experimentally observed in these alloys~\cite{Pd-review}. The physical properties 
that seem to play the essential role in the calculations are: (i) a 
strong host-impurity $d$-$d$ hybridization overwhelming $s$-$d$; (ii) a Pd 
chemical potential poised at a single-particle density-of-states (DOS) peak 
close to the upper $d$-band edge; (iii) a $3d$-impurity 
energy level shifting downwards closer to the host chemical potential by increasing the atomic number 
from Cr towards Ni. The early finding that a simple Anderson impurity model
embodying these three ingredients reproduces correctly at the mean-field level 
the same behavior obtained in more detailed LSDA calculations~\cite{Moriya,Mohn,Zeller-PRB} suggested that no
other ingredient was needed to account for permanent (giant) moments of late
impurities in Pd -- and this might hold in a nanocontact as well.
In order to check whether this conclusion is stable against
quantum fluctuations we therefore start with the standard Anderson impurity model 
\bea
\mathcal{H} &=& 
\sum_{\bk\sigma}\,\epsilon_\bk\,c^\dagger_{\bk\sigma}c^\dagga_{\bk\sigma} + 
\frac{1}{\sqrt{\Omega}}\bigg(V_\bk c^\dagger_{\bk\sigma}d^\dagga_\sigma 
+ H.c.\bigg)\nonumber\\
&& + \epsilon_d\,n_d + \frac{U}{2}\left(n_d-1\right)^2,\label{Ham}
\eea
where $\Omega$ is the volume, $d^\dagger_\sigma$ is the creation operator of an impurity electron 
with spin $\sigma$ and $c^\dagger_{\bk\sigma}$ of a conduction electron with 
the same spin and energy $\epsilon_\bk$. Here $n_d=\sum_\sigma\,d^\dagger_\sigma d^\dagga_\sigma$ 
is the occupation number and $\epsilon_d$ the energy of the impurity level. 
\eqn{Ham} describes a single $d$ orbital hybridized with a single conduction 
band, a deliberately oversimplified situation that nevertheless usually captures the 
essential physics. We consider two extreme cases: (A) a flat conduction electron DOS 
$\rho_A(\epsilon) = 1/2\,\theta\left(1-\epsilon^2\right)$ (all energies are in units of half the conduction bandwidth); 
(B) a substitutional impurity in a one-dimensional chain. In the latter case, it is important to note 
that, while the bulk DOS 
is $\rho_B(\epsilon)=1/(\pi\sqrt{1-\epsilon^2})\,\theta\left(1-\epsilon^2\right)$, the DOS of the hybridizing channel 
is instead $2\sqrt{1-\epsilon^2}/\pi$. The chemical 
potential is taken to be $\mu=0.95$, close to the top of the band, which, in case (B), 
corresponds to a large DOS mimicking the situation of bulk Pd~\cite{Zeller-PRB}.

\begin{figure}
\includegraphics[width=8.7cm]{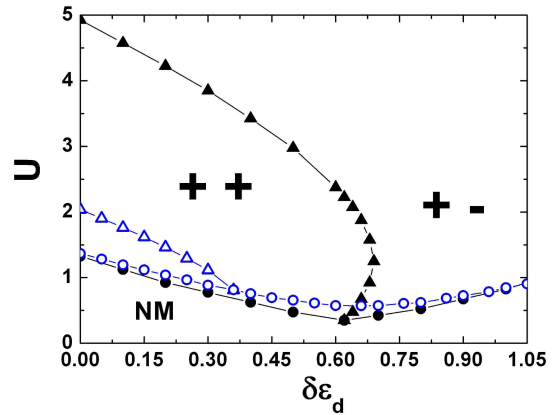}
\caption{\label{fig1} Mean field phase diagram of the model \eqn{Ham} for $\mu=0.95$ and $V_\bk=0.4$. Open and solid 
symbols refer to case (A) and (B), respectively. The circles separate the low $U$ region, where the impurity is non-magnetic at the 
mean field level, from the high $U$ one, where a magnetic mean field solution is stabilized. $++$ denotes the region where the impurity 
is magnetic and co-polarized with the bath, while $+-$ that where is counter-polarized. These two regions are separated by the triangles.}
\end{figure}    

Fig.~\ref{fig1} shows the HF mean field phase diagram as a function 
of $U$ and $\delta \epsilon_d$, the actual impurity level relative to the chemical potential, 
at constant $V_\bk = 0.4$, compatible with a situation of large $d$-$d$ hybridization. We indeed find, confirming
earlier results~\cite{Moriya,Mohn,Zeller-PRB}, that besides wide antiferromagnetically 
impurity-host coupled regions there are (++) regions close to $\delta \epsilon_d=0$ 
where the impurity is magnetic and ferromagnetically co-polarized with the conduction 
bath, again in agreement with LSDA results~\cite{Zeller-PRB}. 
We note in particular that a high DOS at the chemical potential, case (B), does favor the tendency towards magnetism, 
in accord with the Stoner criterion for impurity 
magnetism~\cite{PWA-AIM} (note also the larger region of co-polarization). 
The other element which plays a crucial role 
towards permanent moment formation is the position of the chemical potential 
close to the top of the band. When $\mu$ is shifted towards the center of the 
band, the region of co-polarization shrinks and eventually disappears.  

To study the effect on the permanent moment of quantum fluctuations, we 
choose a value of $U=1.5$ and $\delta \epsilon_d=0.1$ well inside the Hartree-Fock 
ferromagnetically co-polarized (++) region, and we  study the model by NRG.
Surprisingly, we find that NRG reverses the HF result, the co-polarization
disappears and at the fixed point the 
impurity is fully Kondo screened in both cases (A) and (B). 
\begin{figure}
\includegraphics[width=8cm]{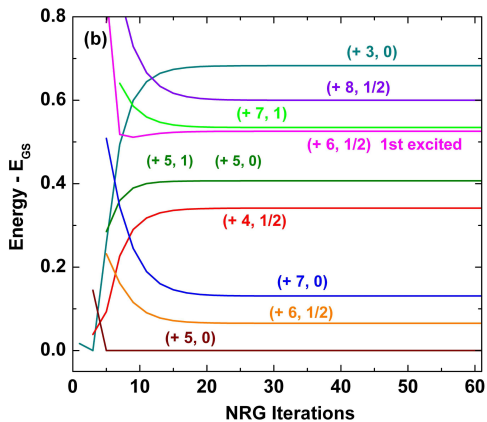}
\caption{\label{fig2} (Color online) NRG flow of the low-energy spectrum for 
Eq.~\eqn{Ham}, case (B), with $U=1.5$, $\delta\epsilon_d=0.1$ and 
odd iterations $N$, i.e. a Wilson chain with even numebr of sites. Each energy level is identified by the charge $Q$ 
with respect to half-filling and the spin $S$. Note the three lowest energy states at large $N$ 
that correspond to the degenerate ground state of a chain with odd number of sites, the impurity site being absorbed by the 
Kondo effect, split by the deviation from particle-hole symmetry.}
\end{figure}    
Fig.~\ref{fig2} 
shows the flow of the many-body low-energy spectrum for case (B) only (the other case is similar) 
versus NRG iteration $N$. $N$ corresponds to a finite temperature of 
order $\Lambda^{-N/2}$ in units of half the conduction bandwidth, where 
$\Lambda$ is the logarithmic discretization parameter~\cite{Krishnamurthy-2} that 
we take equal to three. The asymptotic spectrum is indeed that of a 
Kondo screened impurity below a crossover temperature that can be identified 
with the Kondo temperature $T_K$. This behavior persists on increasing $U$, even though $T_K$ rapidly diminishes.  
For $U\agt 3$, $T_K$ is found to recover the conventional expression 
$T_K\sim \exp\left(-\pi^2 U\rho_0/8\right)$~\cite{Hewson}, where $\rho_0$ is the impurity density of states per spin at the chemical 
potential and in the absence of interaction. For realistic parameters, extracted e.g. 
by LSDA calculations~\cite{Zeller-PRB,nota}, 
the previous expression would imply a substantial $T_K$ of a few hundred K. 
Besides contradicting HF, this large result also excludes the other possible 
explanation for the giant moments in bulk Pd, namely that, because of specific features 
of Pd, the actual coupling is Kondo-like, but with $T_K$ so small 
that, at practical dilutions, the RKKY interaction among the impurities always prevails and 
leads to the ferromagnetism observed in the magnetic alloys.

\begin{figure}
\includegraphics[width=8.7cm]{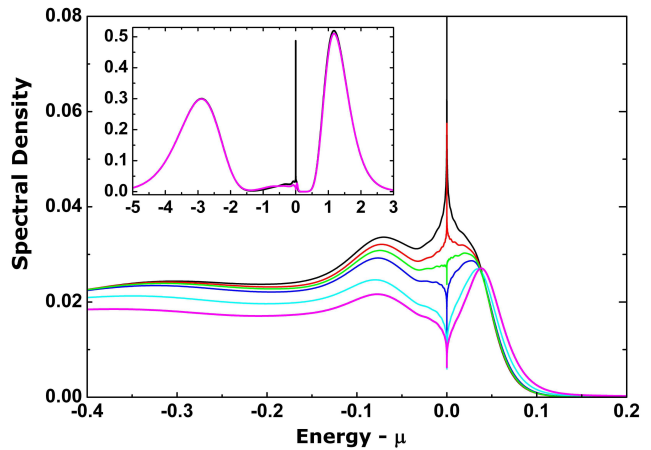}
\caption{\label{fig3} (Color online) Impurity DOS for $U=4$ and different values of the ferromagnetic exchange 
$J$ across the transition, 
$J=0,0.025,0.07,0.1,0.2,0.3$ from top to bottom. 
The large $U$ is used to well separate the low-energy part, shown in the main panel, 
from the high-energy Hubbard bands, shown in the inset
for $J=0.025$ and 0.3.}
\end{figure}

It must be concluded that the Anderson model (\ref{Ham}) is actually insufficient to 
explain the existence of permanent moments in Pd.
Now it is known, in the wide zoo of Kondo models, that the only case where 
a magnetic impurity retains its moment is that of a {\it ferromagnetic} Kondo model~\cite{Hewson}. 
A way to stabilize a permanent moment and to overwhelm 
the effective antiferromagnetic electron-impurity coupling just found by NRG is to 
add an explicit ferromagnetic coupling between the impurity and the conduction 
electrons. Experimental evidence of the important role of such a direct ferromagnetic exchange 
has repeatedly been claimed~\cite{Riegel,Mishra}. Physically, the electron-impurity ferromagnetic exchange  
must be substantial since the host metal is nearly ferromagnetic and the 
impurity wavefunction, much like that of a host-metal atom, extends well over  
neighboring sites. Therefore, we modify the model \eqn{Ham} by 
including a ferromagnetic exchange, of strength $J$, between the impurity 
and the first hybridization shell, namely
\bea
\mathcal{H} \rightarrow \mathcal{H} - J\,\mathbf{S}_c\cdot\mathbf{S}_d, 
\label{Ham1}
\eea
where $\mathbf{S}_d$ is the impurity spin-operator and 
\[
\mathbf{S}_c = \fract{1}{2\sum_\bk \left|V_\bk\right|^2}\,
\sum_{\bk\bkp}\sum_{\alpha\beta}\,V_\bk V_\bkp\,c^\dagger_{\bk\alpha}\,\boldsymbol{\sigma}_{\alpha\beta}\,
c^\dagga_{\bkp\beta},
\]
with $\boldsymbol{\sigma}$ the Pauli matrices. 
Upon repeating the NRG calculation for \eqn{Ham1} we find that 
increasing $J$ 
causes a Berezinskii-Kosterlitz-Thouless phase transition 
at $J=J_*$ from a Kondo screened phase
to an {\it antiscreened} ferromagnetic Kondo phase. 
In the ferromagnetic Kondo state the conduction electrons can be seen as 
scattering off the magnetic impurity and dressing
it up with a spin-cloud whose net result is to restore a fully 
rotationally-invariant state of well defined total spin $S$ 
and zero-point entropy $K_B\ln(2S+1)$. 
The two phases, Kondo screened and antiscreened, also possess different and opposite 
spectroscopic features, as shown by the impurity DOS, $\rho(\epsilon)$ of Fig.~\ref{fig3}. 
On the screened side, $J<J_*$, a Kondo resonance at the chemical potential 
is clearly visible within the Mott-Hubbard gap. On the opposite, antiscreened 
side, $J>J_*$, the impurity DOS develops a cusp-like pseudogap at the chemical potential, $\rho(\epsilon)\sim 1/\ln^2\epsilon$~\cite{Hewson-FM}. 
Unlike the HF approximation, the transition to a permanent moment 
regime is not accompanied by the full disappearance of spectral weight 
within the Mott-Hubbard gap; the pseudogap opens continuously inside 
a broader low-energy feature and is controlled by the scale $J-J_K$, where $J_K$ is the effective antiferromagnetic 
coupling mediated by the hybridization, rather than by $U$. 
From this point of view, the antiferro-ferro Kondo transition just described resembles
the orbital selective Mott transition recently proposed~\cite{DeLeo-Kotliar} 
to explain the paramagnetic-to-antiferromagnetic transition in heavy fermions.

This scenario should describe the spontaneous moment of a magnetic 
impurity embedded in nearly ferromagnetic metal, and equally well the magnetism of a 
bridging impurity in a nanocontact between nearly ferromagnetic leads. 
We note that, while the screened phases is Fermi-liquid like in the strict Nozi\`eres' sense~\cite{Nozieres-JLTP}, 
the antiscreened phase is more properly a {\sl singular} Fermi-liquid~\cite{Coleman&Zarand}. 
At zero temperature and assuming inversion symmetry across the contact, giving rise to well defined even and odd channels,
the zero-bias conductance must recover the Fermi-liquid expression $G=2e^2/h\,\sum_{a}\,\sin^2 \delta_{a}$,  
were $\delta_a$ is the spin-independent difference between the 
scattering phase shifts of the even- and odd-parity scattering channel $a$ at the chemical potential. 
Since the Kondo screened-antiscreened transition is accompanied by a 
jump $\pi/2$ of $\delta_a$ in the channels hybridizing with the magnetic 
orbitals, we should expect a sudden jump of the zero-bias conductance at 
the transition, upward or downward depending on the 
specific symmetry of the impurity and lead orbitals that are involved. 
We also note that, while in the screened regime a magnetic field will, as usual, 
split the Kondo resonance, in the antiscreened phase the magnetic field will
instead fill the pseudogap. Thus, the contact magnetoconductance should behave in 
an opposite way on the two sides of the transition. In addition, in the antiscreened phase  
a non-analytic dependence on the magnetic field and on temperatue should appear~\cite{Coleman&Zarand,Hewson-FM}.

Experimental observations of a ferromagnetic Kondo effect at nanocontacts --
even if probably already present -- are not obvious to pinpoint in existing data.  
Transition metal impurities such as Fe or Co, bare or embedded in molecules
such as porphyrins or phtalocyanines, contacted by a nearly ferromagnetic 
metal such as Pd, could represent a starting point. Bare Pd break junctions
could develop a local moment at the contact~\cite{Delin-PRL}, which might lead to a 
ferro Kondo zero-bias anomaly.
Interestingly, Kondo-liked anomalies are apparently being found in pure Fe, Co and Ni
break junctions\cite{untiedt}. Their relatively small width supports their 
Kondo origin. However, in view of the many orbital channels that are involved, these anomalies could equally well 
indicate regular antiferro Kondo or, as previously discussed, 
a ferro Kondo. 
Pt or Ir break junctions would also be very interesting systems to look for zero-bias
anomalies, owing to the local moment 
which is expected to form in the chain-like 
nanocontact~\cite{smit}. Even though the susceptibility of Pt and Ir is not
as Stoner-enhanced as that of Pd, it is plausible that there should be enough
intersite ferromagnetic exchange in these metals to lead to a ferromagnetic
coupling of the leads to the bridging chain. It should be noted however 
that strong spin-orbit coupling should also give rise here to a colossal~\cite{Mokrousov} 
or giant magnetic anisotropy~\cite{Delin-Nature,ferrer}, an interesting complication not 
yet described by the present treatment.

This work was sponsored by MIUR PRIN/Cofin Contract No. 2006022847 as well 
as by INFM/CNR ``Iniziativa transversale calcolo parallelo''. The research 
environment provided by the independent ESF project CNR-FANAS-AFRI was also 
useful. We gratefully acknowledge correspondence with C. Untiedt.


\begin{thebibliography}{27}
\bibitem[{not()}]{nota}
\bibinfo{note}{$U$ can be estimated in LSDA by the spin splitting of the
  impurity DOS, while $\rho_0$ can be approximated by the value of the impurity
  DOS per spin and orbital in LDA, i.e. preventing spin-symmetry breaking. For
  instance, in Ref.~\cite{Zeller-PRB} these values are found to be $U\sim 4$~eV
  and $\rho\simeq 1$~States/eV.}
\end{thebibliography}

\end{document}